\documentclass[preprint,12pt]{elsarticle}
\usepackage{graphicx}
\usepackage{subeqn}
\usepackage{amssymb}
\journal{Physica A}

\usepackage{ulem} 

\usepackage[usenames]{color}

\newcommand{\bk}{{\bf k}}
\newcommand{\bl}{{\bf l}}

\newcommand{\cE}{{\mathcal E}}

\newcommand{\cH}{{\mathcal H}}

\begin{document}

\author[IFIN]{Drago\c s-Victor Anghel}\corref{CA}
\ead{dragos@theory.nipne.ro}

\author[IFIN,UniBuc]{George Alexandru Nemnes}
\ead{nemnes@theory.nipne.ro}

\cortext[CA]{Corresponding author}

\address[IFIN]{Institutul National de Fizica si Inginerie Nucleara Horia Hulubei, P.O.BOX MG-6, RO-077125, M\u agurele, Jud. Ilfov, Romania}

\address[UniBuc]{University of Bucharest, Faculty of Physics, ``Materials and Devices for Electronics and Optoelectronics'' Research Center, 077125 M\u agurele-Ilfov, Romania}

\title{The application of the fractional exclusion statistics to the BCS theory--a redefinition of the quasiparticle energies}

\date{\today}

\begin{abstract}
The effective energy of a superconductor $E_{eff}(T)$ at temperature $T$ is defined as the difference between the total energy at temperature $T$ and the total energy at 0~K. We call the energy of the condensate, $\cE_c$, the difference between $E_{eff}$ and the sum of the quasiparticle energies $E_{qp}$.
$\cE_c$, $E_{qp}$, as well as the BCS quasiparticle energy $\epsilon$ are positive and depend on the gap energy $\Delta$, which, in turn, depends on the populations of the quasiparticle states (equivalently, they depend on $T$). So from the energy point of view the superconductor is a Fermi liquid of non-ideal quasiparticles.

We show that the choice of quasiparticles is not unique, but there is an infinite range of possibilities.
Some of these possibilities have been explored in the context of the fractional exclusion statistics (FES), which is a general method of describing interacting particle systems as ideal gases.
We apply FES here and transform the Fermi liquid of BCS excitations into an ideal gas by redefining the quasiparticle energies.
The new FES quasiparticles exhibit the same energy gap as the BCS quasiparticles, but a different DOS, which is finite at any quasiparticle energy.

We also discuss the effect of the remnant electron-electron interaction (electron-electron interaction beyond the BCS pairing model) and show that this can stabilize the BCS condensate, increasing the critical temperature.
\end{abstract}

\maketitle

\section{Introduction} \label{intro}

We divide the energy of a BCS superconductor \cite{PhysRev.108.1175.1957.Bardeen} into three parts: the ground-state energy $E_{gs}$, the condensate energy $\cE_c$, and the energy of the quasiparticles $E_{qp}$.
$E_{gs}$ is a constant and represents the total energy of the superconductor at temperature $T=0$, $E_{qp}$ is the sum of the excitations' quasiparticle energies, whereas the condensate energy is the difference $\cE_c \equiv E - E_{gs} - E_{qp}$. $\cE_c$ vanishes at $T=0$ and increases monotonically with $T$, reaching its highest value at the critical temperature $T_c$, where the superconducting state disappears.
Effectively, the energy of the system--after removing the constant term $E_{gs}$--is $E_{eff} \equiv \cE_c + E_{qp}$. Due to the fact that both, $\cE_c$ and the BCS quasiparticle energies $\epsilon$, depend on the populations of the quasiparticle states $\{n_\epsilon\}$, $E_{eff}$ represents the energy of a Fermi liquid (FL) \cite{JETP.3.920.1956.Landau, PinesNozieres:book} and $\epsilon\equiv \partial E_{eff}/\partial n_\epsilon$.

In the context of fractional exclusion statistics (FES) \cite{PhysRevLett.67.937.1991.Haldane, PhysRevLett.73.922.1994.Wu, PhysRevLett.73.2150.1994.Isakov} it has been shown that the quasiparticle energies may be redefined (see e.g. Refs.~\cite{PhysRevLett.73.3331.1994.Murthy, PhysRevLett.74.3912.1995.Sen, NewDevIntSys.1995.Bernard, JPhysB33.3895.2000.Bhaduri, PhysRevLett.86.2930.2001.Hansson, PhysRevE.85.011144.2012.Liu, JStatMechThExp.P01034.2015.Sharma, JStatMech.P04018.2013.Gundlach, CommunTheorPhys.62.2014.81.Fang, PhysLettA.372.5745.2008.Anghel, PhysScr.2012.014079.2012.Anghel, PhysRevE.88.042150.2013.Anghel}). There is an infinite range of possibilities in which one can redistribute the energy of the system among the quasiparticle states.
Moreover, if the choice is made such that the total (or the effective) energy of the system is equal to the sum of the quasiparticle energies, then one obtains a description of the system in terms of an ideal FES gas \cite{PhysScr.2012.014079.2012.Anghel}. All the choices of quasiparticle energies must lead to thermodynamically equivalent descriptions, in the sense that the populations of the quasiparticle states and all the macroscopic thermodynamic quantities should not depend on the chosen description \cite{PhysScr.2012.014079.2012.Anghel, PhysRevE.88.042150.2013.Anghel}. 
We exemplify the procedure by redefining the quasiparticle in such a way that $E_{eff}$ becomes the sum of the new quasiparticle energies $\tilde\epsilon$. This relation holds for any quasiparticle levels populations, so the system obtained is an ideal gas.
In our example the quasiparticle energies $\tilde\epsilon$ exhibits the same energy gap $\Delta$ as the BCS quasiparticles, but the density of states (DOS) $\tilde\sigma(\tilde\epsilon)$ is finite over the whole spectrum (including at $\tilde\epsilon=\Delta$). 
%

We also extend the BCS model by including an extra interaction between the electrons as a perturbation to the initial pairing Hamiltonian. This leads to an interaction term between the quasiparticles which modifies the energy gap and the quasiparticle energies.
The gap equation cannot be satisfied anymore for $\Delta = 0$ at any temperature, so, in the first order of perturbation, the extra interaction does not allow the superconducting phase to be destroyed.

The paper is organized as follows. In the next subsection we introduce the notations and the basic concepts of the BCS theory. Then, in Section \ref{sec_energy}, we write the effective energy of the system as the energy of a Fermi liquid (FL), with the BCS quasiparticle energies equal to the Landau's quasiparticle energy of the FL. The FES description is presented in Section \ref{sec_FES0}, where we introduce the FES quasiparticle energies, the FES parameters, and we write the FES equations for the population. We also show that the FES and and FL descriptions are physically equivalent. In Section~\ref{sec_form_ext} we extend the BCS model by introducing the interaction between the quasiparticles. In Section~\ref{conclusions} we present the conclusions.

\subsection{The basics of the theory of superconductivity}
\label{basics}

Let us specify notations and the basic ideas of the BCS theory, following mainly Refs. \cite{Tinkham:book, PhysRev.108.1175.1957.Bardeen}.
We denote the single-particle states of the electrons in the superconductor by $|\bk,s\rangle$ and its time reversed state by  $|-\bk,-s\rangle$; $s$ is the spin and $\bk$ represents the rest of single-particle quantum numbers that specify the state. Concretely, in the following we shall consider that $\bk$ is the free electron wavevector.
The electrons creation and annihilation operators are $c^\dagger_{\bk, s}$ and $c_{\bk,s}$, respectively, and the BCS pairing Hamiltonian is
\begin{equation}
  \cH_{BCS} = \sum_{\bk\sigma}\epsilon^{(0)}_\bk n_{\bk\sigma} + \sum_{\bk\bl} V_{\bk\bl} c^\dagger_{\bk\uparrow} c^\dagger_{-\bk\downarrow} c_{-\bl\downarrow} c_{\bl\uparrow},  \label{def_H_BCS}
\end{equation}
where $\epsilon^{(0)}_\bk$ are the energies of the non-interacting single-particle states and $V_{\bk\bl}$ are the matrix elements of the attractive effective interaction potential.
The ground state will be denoted by $|BCS\rangle_0$. 
The Hamiltonian (\ref{def_H_BCS}) is diagonalized by the Bogoliubov transformations, by writing
%
  $c_{-\bk\downarrow} c_{\bk\uparrow} \equiv b_\bk + (c_{-\bk\downarrow} c_{\bk\uparrow} - b_\bk)$,
%
where $b_\bk = \langle c_{-\bk\downarrow} c_{\bk\uparrow} \rangle$, and assuming that $c_{-\bk\downarrow} c_{\bk\uparrow} - b_\bk$ is small ($\langle \cdot \rangle$ is the average). Then $\cH_{BCS}$ is expanded in terms of $c_{-\bk\downarrow} c_{\bk\uparrow} - b_\bk$ and keeping only the first order we get
\begin{equation}
  \cH = \sum_{\bk\sigma}\epsilon^{(0)}_\bk n_{\bk\sigma} + \sum_{\bk\bl} V_{\bk\bl} ( c^\dagger_{\bk\uparrow} c^\dagger_{-\bk\downarrow} b_\bl + b^*_\bk c_{-\bl\downarrow} c_{\bl\uparrow} - b^*_\bk b_\bl). \label{def_H_BCS_approx}
\end{equation}
We define the model Hamiltonian $\cH_M = \cH - \mu N$, which can be diagonalized to become \cite{Tinkham:book}
\begin{equation}
  \cH_M = \sum_\bk(\xi_\bk-\epsilon_\bk+\Delta_\bk b_\bk^*) + \sum_\bk\epsilon_\bk(\gamma^\dagger_{\bk 0}\gamma_{\bk 0} + \gamma^\dagger_{\bk 1}\gamma_{\bk 1}) , \label{HM_BCS}
\end{equation}
where $\xi_\bk\equiv \epsilon^{(0)}_\bk - \mu$, $\epsilon_\bk \equiv\sqrt{\xi_\bk^2+\Delta_\bk^2}$ and $\Delta_\bk$ is the energy gap,
\begin{eqnarray}
  \Delta_\bk &=& -\sum_l V_{\bk\bl} \langle c_{-\bl\downarrow} c_{\bl\uparrow}\rangle . \label{def_Delta0}
\end{eqnarray}
The operators $\gamma^\dagger_{\bk i}$ and $\gamma_{\bk i}$ ($i=0,1$) are quasiparticle creation and annihilation operators, respectively, and are defined by the relations
\begin{subequations} \label{def_cs}
\begin{eqnarray}
  c_{\bk\uparrow} &=& u^*_\bk \gamma_{\bk 0} + v_\bk \gamma^\dagger_{\bk 1} , \label{def_c1} \\
  c^\dagger_{\bk\uparrow} &=& u_\bk \gamma^\dagger_{\bk 0} + v^*_\bk \gamma_{\bk 1} , \label{def_c2} \\
  c^\dagger_{-\bk\downarrow} &=& -v^*_\bk \gamma_{\bk 0} + u_\bk \gamma^\dagger_{\bk 1} , \label{def_c3} \\
  c_{-\bk\downarrow} &=& -v_\bk \gamma^\dagger_{\bk 0} + u^*_\bk \gamma_{\bk 1} . \label{def_c4}
\end{eqnarray}
\end{subequations}
The coefficients $u_\bk$ and $v_\bk$ satisfy the relation
\begin{equation}
  |v_\bk|^2 = 1 - |u_\bk|^2 = \frac{1}{2} \left(1 - \frac{\xi_\bk}{\epsilon_\bk}\right) . \label{def_uv}
\end{equation}

We work with the typical assumption that $V_{\bk \bl}\equiv V$ for any $\bk$ and $\bl$ in an energy shell of $2\hbar\omega_c$ around the Fermi energy, and $V_{\bk \bl}=0$ outside this shell.
Using Eqs. (\ref{def_Delta0}) and (\ref{def_uv}) and denoting the quasiparticle populations by $n_{\bk i} \equiv \langle \gamma^\dagger_{\bk i}\gamma_{\bk i} \rangle$, we obtain the self-consistency relation for $\Delta$,
\begin{equation}
  1 = \frac{V}{2} \sum_\bk \frac{1 - n_{\bk 0} - n_{\bk 1}}{\epsilon_\bk} . \label{def_Delta2}
\end{equation}
Working in the quasicontinuous limit we transform the summations into integrals. Changing the integration variable from $\bk$ to $\xi$, Eq. (\ref{def_Delta2}) becomes
\begin{eqnarray}
  \frac{2}{V} &=& \int_{-\hbar\omega_c}^{\hbar\omega_c} \frac{1 - n_{\xi 0} - n_{\xi 1}}{\sqrt{\xi^2+\Delta^2}} \sigma_0\,d\xi , \label{Eq_int_Delta1}
\end{eqnarray}
where we assumed that the DOS over the $\xi$ axis $\sigma_0$ is constant.
If we set $n_{0\xi} = n_{1\xi} = 0$, we obtain in the weak-coupling limit $\sigma_0 V\ll 1$ the energy gap at zero temperature $\Delta_0 = 2\hbar\omega_c\exp[-1/(\sigma_0V)]$.
The critical temperature $T_c$ may be calculated from Eq. (\ref{Eq_int_Delta1}) by setting $\Delta = 0$ and we get $k_BT_c = A \hbar\omega_c e^{-1/(\sigma_0 V)}$, where $A\approx 1.1339$ \cite{Tinkham:book}.

The number of excitations at temperature $T$ is defined as:
\begin{equation}
N_{ex}(T) = \sum_{\bk} (n_{\bk 0} + n_{\bk 1}) = 2\sigma_0\int_\Delta^\infty
	\frac{\epsilon}{\sqrt{\epsilon^2-\Delta^2}} (n_{\bk 0} + n_{\bk 1}) d\epsilon .
\end{equation}
From Eq. (\ref{Eq_int_Delta1}) we find an expression for the total number of excitations at critical temperature $N_{ex}(T_c)$:
\begin{eqnarray}
  N_{ex}(T_c) &=& (4 \ln 2) \sigma_0 k_BT_c = (4 \ln 2) \sigma_0 A \hbar\omega_c e^{-1/(\sigma_0 V)} .
  \label{Nex_Tc}
\end{eqnarray}
In the low temperature limit we find the asymptotic behavior
\begin{equation}
  N_{ex}(T\to 0) \approx 4 \sigma_0 k_BT \sqrt{\frac{y_0}{2}} e^{-y_0} \int_0^\infty \frac{e^{-z} \, dz}{\sqrt{z}}
  = 2\sqrt{2\pi} \sigma_0 \sqrt{k_BT \Delta_0} e^{-\beta\Delta_0} .
  \label{Nex_T0}
\end{equation}
In Eq. (\ref{Nex_T0}) we shall always consider that $\sigma_0 k_BT > 1$, i.e. the inter-level energy spacing is always smaller than the thermal energy $k_BT$.
Under this assumption we have $2\sqrt{2\pi} \sigma_0 \sqrt{k_BT \Delta_0} \gg 1$ for any \textit{physically significant} temperature.

To find a low temperature asymptotic expression for $\Delta$ we write Eq. (\ref{Eq_int_Delta1}) in the low temperature limit as
\begin{eqnarray}
  \frac{2}{V} &\stackrel{T\to 0}{\sim}& 2\sigma_0 \ln\left[\frac{\hbar\omega_c}{\Delta} + \sqrt{\left(\frac{\hbar\omega_c}{\Delta}\right)^2+1}\right] - 2 \sqrt{2\pi}\sigma_0 e^{-\beta\Delta} , \label{Eq_Delta0_1}
\end{eqnarray}
from where we obtain
\begin{eqnarray}
  \Delta_0 - \Delta \stackrel{T\to 0}{\sim} \sqrt{2\pi} \Delta_0 e^{-\beta\Delta_0}
  \label{Eq_Delta0_2}
\end{eqnarray}

\section{The energy of the system} \label{sec_energy}

From Eqs. (\ref{def_H_BCS_approx}) and (\ref{HM_BCS}) we write the energy of the system as
\begin{equation}
  \cH = E_0 + \cH_{\rm qp} = E_0 + \sum_\bk\epsilon_\bk (\gamma^\dagger_{\bk 0}\gamma_{\bk 0} + \gamma^\dagger_{\bk 1}\gamma_{\bk 1}) , \label{HM_BCS_eff}
\end{equation}
where
\begin{equation}
  E_0 = \mu N + \sum_\bk(\xi_\bk-\epsilon_\bk+\Delta b_\bk^*)
  \equiv \mu N + \sum_\bk (\xi_\bk - \epsilon_\bk) + \frac{\Delta^2}{V} \label{def_E0}
\end{equation}
and quasiparticles energy is
\begin{equation}
  E_{qp} \equiv \sigma_0 \int_{-\hbar\omega_c}^{\hbar\omega_c} ( n_{\xi 0} + n_{\xi 1}) \epsilon(\xi) \, d\xi \ge 0 . \label{def_cHpqp}
\end{equation}
We denote by $E_{0N}$ the energy of a free electron gas with the same number of particles $N$, at zero temperature, and we define the quantity \cite{Tinkham:book}
\begin{eqnarray}
  \cE_0 &\equiv& E_0 - E_{0N} = \sum_{k> k_\mu} (\xi_\bk - \epsilon_\bk) + \sum_{k \le k_\mu} (- \xi_\bk - \epsilon_\bk) + \frac{\Delta^2}{V} \nonumber \\
  &=& - \frac{\sigma_0\Delta^2}{2} \left[ 1 + 2 \ln\left(\frac{\Delta_0}{\Delta}\right) \right] \le 0 . \label{estim_Us_Un}
\end{eqnarray}
In the absence of excitations ($T=0$), $\Delta = \Delta_0$ and $\cE_0$ takes its minimum value $\cE_{0g} \equiv \cE_0(T=0) = - \sigma_0\Delta_0^2/2$.
%
The minimum energy of the superconductor and the condensate energy (which was specified in the Introduction) are
\begin{subequations} \label{defs_Egs_Ec}
\begin{eqnarray}
  E_{gs} &\equiv& E_{0N} + \cE_{0g} \equiv E_{0N} - \sigma_0\Delta_0^2/2 \quad {\rm and} \label{def_Egs} \\
  \cE_c &\equiv& \cE_0 - \cE_{0,g} = \frac{\sigma_0 \Delta_0^2}{2} \left\{ 1 - \left( \frac{\Delta}{\Delta_0} \right)^2 \left[ 1 - 2\ln\left( \frac{\Delta}{\Delta_0} \right) \right] \right\} , \label{def_Ec}
\end{eqnarray}
\end{subequations}
respectively.
Using Eqs. (\ref{def_cHpqp}) and (\ref{def_Ec}) and the definition of $E_{eff} \equiv \cE_c + E_{qp}$, we introduce the effective Hamiltonian of the superconductor
\begin{eqnarray}
  \cH_{eff} &\equiv& \cE_c + \sum_{\bk} \sum_{i=0,1} \epsilon_{\bk} \gamma^\dagger_{\bk i}\gamma_{\bk i} \label{HM_BCS_eff2}
\end{eqnarray}
and $E_{eff} \equiv \langle \cH_{eff} \rangle$.
In Eq. (\ref{HM_BCS_eff2}) both, $\Delta$ and $\epsilon_{\bk}$, depend on the of set of populations $n_{\bk i}$. The effective energy is that of an interacting system of fermions, described in the Fermi liquid theory (FLT):
\begin{eqnarray}
  E_{eff} (\{n\}) &=& \cE_c (\{n\}) + \sum_{\bk, i} n_{\bk i} \epsilon_{\bk}(\{n\}) , \label{cE_dep_nki}
\end{eqnarray}
where by $\{n\}$ we explicitly specified the dependence of $\cE$ and $\epsilon_{\bk}$ on the whole set of populations.

To calculate the temperature dependence of the population, let us write
\begin{subequations} \label{dcE0_dEqp_dNki}
\begin{equation}
  \partial E_{eff}/\partial n_{\bk i} = \partial \cE_c/\partial n_{\bk i} + \partial E_{qp}/\partial n_{\bk i}, \label{dEeff_dNki}
\end{equation}
where
\begin{eqnarray}
  \frac{\partial \cE_c}{\partial n_{\bk i}} &=&   \frac{\partial\cE_c}{\partial\Delta} \frac{\partial\Delta}{\partial n_{\bk i}} \equiv \epsilon_{\bk i}' \quad {\rm and} \label{dcE0_dNki} \\
  \frac{\partial E_{qp}}{\partial n_{\bk i}} &=& \epsilon_\bk +  \frac{\partial E_{qp}}{\partial\Delta} \frac{\partial\Delta}{\partial n_{\bk i}} \equiv \epsilon_\bk + \epsilon_{\bk i}'' . \label{dEqp_dNki}
\end{eqnarray}
\end{subequations}
From Eqs. (\ref{def_cHpqp}), (\ref{def_Ec}), and (\ref{def_Delta2}) we obtain
\begin{subequations} \label{dcE_dHqp_dDelta}
\begin{eqnarray}
  \frac{\partial \cE_c}{\partial \Delta} &=&  \frac{2\Delta}{V} - \Delta  \sigma_0 \int_{-\hbar\omega_c}^{\hbar\omega_c} \frac{d\xi}{\epsilon_\xi}  , \label{dcE_dDelta} \\
  \frac{\partial E_{qp}}{\partial \Delta} &=& - \frac{2 \Delta}{V} + \Delta \sigma_0 \int_{-\hbar\omega_c}^{\hbar\omega_c} \frac{d\xi}{\epsilon_\xi}  ,\label{dEqp_dDelta}
\end{eqnarray}
\end{subequations}
so $\partial E/\partial\Delta = \partial (\cE_c+E_{qp})/\partial \Delta = 0$. 
Then the quasiparticle energy defined in the FLT sense is exactly the BCS quasiparticle energy,
\begin{equation}
  \epsilon_\bk = \frac{\partial E_{eff}}{\partial n_{\bk i}} , \label{dE_dNbk_tot}
\end{equation}
which further leads to the typical equilibrium BCS distribution
\begin{equation}
  n_{\bk i} = \frac{1}{e^{\beta \epsilon_\bk} + 1} , \label{pop_epsilon}
\end{equation}
where $\beta = (k_BT)^{-1}$ and $T$ is the temperature.

\section{The implementation of the fractional exclusion statistics} \label{sec_FES0}

\subsection{The quasiparticle energies and the density of states} \label{sec_FES}

\begin{figure}[t]
  \centering
  \includegraphics[width=4cm]{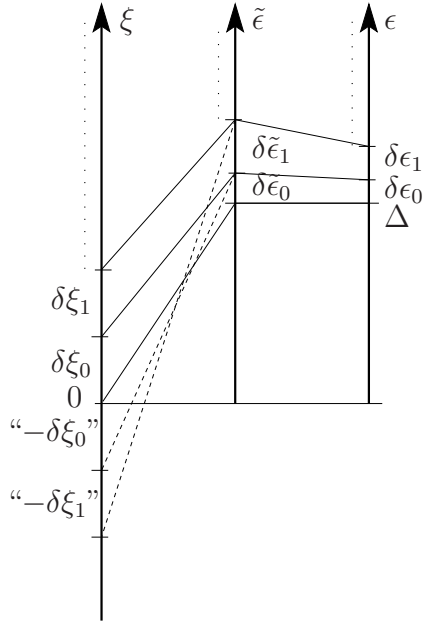}
  \put(-100,225){$\xi$}
  \put(-126,119){$\delta\xi_1$}
  \put(-126,95){$\delta\xi_0$}
  \put(-120,82){0}
  \put(-143,42){``$-\delta\xi_1$''}
  \put(-143,69){``$-\delta\xi_0$''}
  \put(-50,225){$\tilde\epsilon$}
  \put(-50,176){$\delta\tilde\epsilon_1$}
  \put(-50,162){$\delta\tilde\epsilon_0$}
  \put(0,225){$\epsilon$}
  \put(0,173){$\delta\epsilon_1$}
  \put(0,161){$\delta\epsilon_0$}
  \put(0,150){$\Delta$}
  \caption{The FES quasiparticle species represented as intervals on the FES quasiparticle energy axis (middle) $\delta\tilde\epsilon_0, \delta\tilde\epsilon_1, \ldots$. Each species corresponds to two intervals (symmetric with respect to the origin, i.e. the Fermi surface) on the $\xi$ axis (left) and one interval on the $\epsilon$ axis (right).
  The relations between $\xi$, $\epsilon$, and $\tilde\epsilon$ are given in Eqs. (\ref{inv_def_qpen}).
  Both, $\epsilon$ and $\tilde\epsilon$, have the same energy gap $\Delta$.}
  \label{mapping}
\end{figure}

\begin{figure}[t]
  \centering
  \includegraphics[width=8cm]{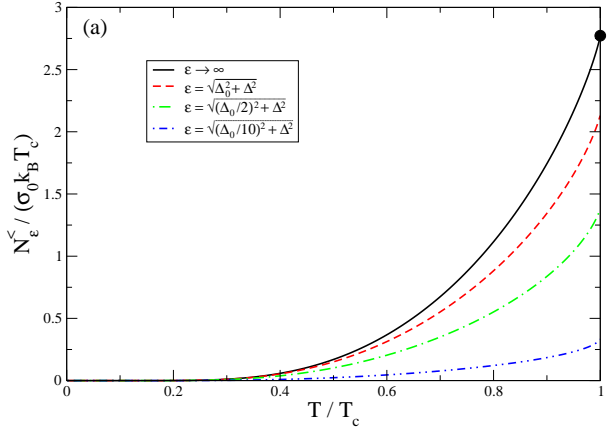}
  \caption{(Color online) The scaled number of excitations of energy below $\epsilon$, $N_{\epsilon}^</(\sigma_0 k_B T_c)$, vs. the scaled temperature $T/T_c$ for several values $\epsilon$, where $N_{\infty}^< \equiv N_{ex}$. The dot marks the value for $N_{ex}(T_c)/(\sigma_0 k_B T_c)=4\log(2)$ as indicated in Eq.~(\ref{Nex_Tc}).
  }
  \label{Nex_plot}
\end{figure}

\begin{figure}[t]
  \centering
  \includegraphics[width=8cm]{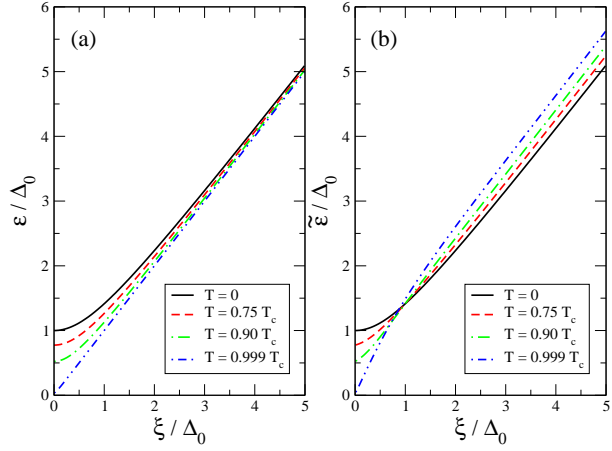}
  \caption{(Color online) Quasiparticle energies vs. $\xi$, for three different temperatures $T=0,\;0.75,\;0.9,\;0.999$ $T_c$: (a) $\epsilon(\xi)$ and (b) $\tilde\epsilon(\xi)$. At $\xi=0$ we have $\tilde\epsilon=\epsilon=\Delta(T)$ for any $T$.
  }
  \label{qp_energies}
\end{figure}

The BCS Hamiltonian (\ref{HM_BCS_eff2}) describes a FLT system of interacting fermions \cite{PinesNozieres:book}.
We want to transform this into an ideal gas Hamiltonian using a method similar to that outlined in Refs. \cite{PhysLettA.372.5745.2008.Anghel, PhysScr.2012.014079.2012.Anghel, PhysRevE.88.042150.2013.Anghel}.
For this we define \textit{new quasiparticle energies} $\tilde\epsilon$, such that
\begin{eqnarray}
  E_{eff} &=& \sum_{\tilde\epsilon, i} n_{\tilde\epsilon i} \tilde \epsilon .  \label{def_E_qpen}
\end{eqnarray}
%
The ideal gas thus obtained obeys FES \cite{PhysRevLett.67.937.1991.Haldane, PhysRevLett.73.922.1994.Wu} and a schematic depiction of the quasiparticle species is presented in Fig. \ref{mapping}.
The new quasiparticle energies are  related to $\epsilon$ by the relation
\begin{subequations} \label{defs_qpen_Neps_cE}
\begin{eqnarray}
  \tilde \epsilon(\epsilon) &\equiv& \epsilon + \frac{2\cE_c}{N_{ex}(N_{ex}+1)} N_{\epsilon}^<
  \approx \epsilon + \frac{2\cE_c}{N_{ex}^2} N_{\epsilon}^< \nonumber \\
  &\equiv& \xi + (\sqrt{\epsilon^2 - \Delta^2} - \xi) + \frac{2\cE_c}{N_{ex}^2} N_{\epsilon}^<, \label{def_qpen}
\end{eqnarray}
where $N_{\epsilon}^<$ is the number of excitations of energy below $\epsilon$, namely
\begin{equation}
  N_{\epsilon}^< \equiv \sigma_0 \int\limits_{-\sqrt{\epsilon^2 - \Delta^2}}^{\sqrt{\epsilon^2 - \Delta^2}} (n_{\xi 0} + n_{\xi 1}) \, d\xi
  = 2 \sigma_0 \int_\Delta^\epsilon (n_{\epsilon' 0} + n_{\epsilon' 1}) \frac{\epsilon' \, d\epsilon'}{\sqrt{(\epsilon')^2 - \Delta^2}} . \label{def_N_eps}
\end{equation}
\end{subequations}
The temperature dependence of $N_{\epsilon}^<$ for several values of $\epsilon(\xi)$ is shown in Fig.~\ref{Nex_plot}.
Since $\cE_c \ge 0$, Eq. (\ref{def_qpen}) implies that $\tilde\epsilon(\epsilon_1) \le \tilde\epsilon(\epsilon_2)$ if $\epsilon_1 \le \epsilon_2$.
Disregarding the degeneracy of the quasiparticle energy levels, Eq. (\ref{def_qpen}) defines a bijective transformation and therefore we may also use the notation $N_{\tilde\epsilon(\epsilon)}^< \equiv N_{\tilde\epsilon}^<$.
The new definition of quasiparticle energy satisfies Eq. (\ref{def_E_qpen}).
Inverting Eq. (\ref{def_qpen}) we get
\begin{equation}
  \epsilon (\tilde\epsilon) = \tilde\epsilon - \frac{2\cE_c}{N_{ex}^2} N_{\epsilon}^< \quad {\rm and} \quad 
  \xi (\tilde\epsilon) = \pm \sqrt{\left(\tilde\epsilon - \frac{2\cE_c}{N_{ex}^2} N_\epsilon^<\right)^2 - \Delta^2} . \label{inv_def_qpen}
\end{equation}
Also from Eq. (\ref{def_qpen}) we see that $\tilde\epsilon(\epsilon = \Delta) = \Delta$, so the FES quasiparticle energies have the same gap as the BCS quasiparticles. For high energies, $\tilde\epsilon(\epsilon \gg \Delta) \approx \epsilon + 2\cE_c/N_{ex} \approx \xi + 2\cE_c/N_{ex}$.
The quasiparticle energies $\tilde\epsilon$ and $\epsilon$ are represented as functions of $\xi$
in Fig.~\ref{qp_energies}.

\begin{figure}[t]
  \centering
  \includegraphics[width=8cm]{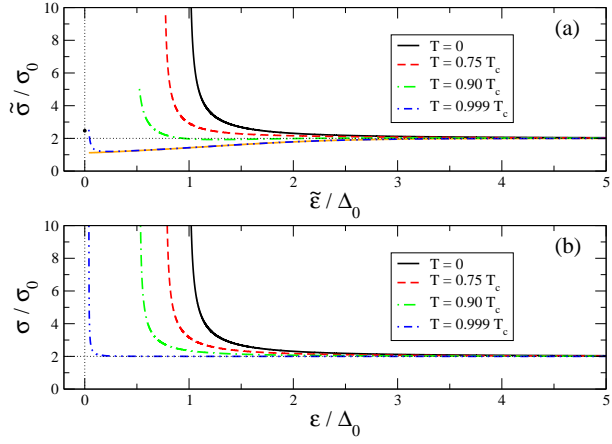}
  \caption{(Color online) Quasiparticle density of states, scaled by $\sigma_0$, for the temperatures $T=0,\;0.75,\;0.9,\;0.999$ $T_c$: (a) $\tilde\sigma/\sigma_0$ vs. $\tilde\epsilon/\Delta_0$ and (b) $\sigma/\sigma_0$ vs. $\epsilon/\Delta_0$.
  While $\lim_{\epsilon\searrow\Delta}\sigma(\epsilon)=\infty$ at any temperature, $\tilde\sigma(\tilde\epsilon)$ is finite at any $T>0$.
  In (a), the dot at $\tilde\epsilon = 0$ and $\tilde\sigma/\sigma_0 \approx 2.47$ corresponds to Eq.~(\ref{tilsig_tileps_Tc_disc}) and the lowest ``color'' continuous line is $\tilde\sigma(\tilde\epsilon,T_c)$ (Eq. \ref{tilsig_tileps_Tc}).
  }
  \label{qp_dos}
\end{figure}

The DOS along the $\tilde\epsilon$ axis (which we call the FES DOS) is $\tilde\sigma(\tilde\epsilon, T)$ and may be calculated from Eqs. (\ref{defs_qpen_Neps_cE}) using the identity $\tilde\sigma(\tilde\epsilon)/ \sigma(\epsilon) =  (d\tilde\epsilon/d\epsilon)^{-1}$:
\begin{eqnarray}
  \tilde\sigma(\tilde\epsilon, T) &=& \frac{2\sigma_0 \epsilon(\tilde\epsilon)}{\sqrt{\epsilon^2(\tilde\epsilon) - \Delta^2} + \epsilon(\tilde\epsilon) [4\cE_c/N_{ex}^2] \sigma_0 (n_{0 \epsilon} + n_{1\epsilon})} , \label{def_tilsig}
\end{eqnarray}
where $\tilde\epsilon$ and $\epsilon$ are related by Eqs. (\ref{def_qpen}) and (\ref{inv_def_qpen}). When $\tilde\epsilon = \Delta$ we have
\begin{eqnarray}
  \tilde\sigma(\Delta, T) &=& \frac{N_{ex}^2}{2\cE_c (n_{\Delta 0} + n_{\Delta 1})} 
  = \frac{N_{ex}^2}{n_{\Delta 0} + n_{\Delta 1}} \frac{1}{\sigma_0 \Delta_0^2} \nonumber \\
  && \times \left\{ 1 - \left( \frac{\Delta}{\Delta_0} \right)^2 \left[ 1 - 2\ln\left( \frac{\Delta}{\Delta_0} \right) \right] \right\}^{-1}
  . \label{def_tilsig_Delta}
\end{eqnarray}
and we observe that $\tilde\sigma(\Delta, T)<\infty$ for any $T>0$. In the low temperature limit,
\begin{eqnarray}
  \tilde\sigma(\Delta, T\to 0) &\sim& \frac{N_{ex}^2}{n_{\Delta 0} + n_{\Delta 1}} \frac{1}{2\sigma_0 (\Delta_0 - \Delta)^2} \approx \sigma_0 \frac{k_BT}{\Delta_0} e^{\frac{\Delta_0}{k_BT}}, \label{def_tilsig_Delta_T0}
\end{eqnarray}
where for the last expression we used Eqs. (\ref{Nex_T0}) and (\ref{Eq_Delta0_2}).
At the critical temperature
\begin{eqnarray}
  \tilde\epsilon(\xi \equiv\epsilon, T_c) &=& \xi + \frac{1}{2 A^2 \ln^2 2} \int_0^\xi \frac{d\xi'}{e^{\xi'/(k_BT_c)}+1} \label{tileps_Tc}
\end{eqnarray}
%
and
\begin{equation}
  \tilde\sigma(\tilde\epsilon, T_c) = 2\sigma_0 \left\{ 1 + \frac{1}{A^2 \ln^2 2 [e^{\xi(\tilde\epsilon) /(k_BT_c)} + 1]} \right\}^{-1} , \label{tilsig_tileps_Tc}
\end{equation}
where in Eq. (\ref{tilsig_tileps_Tc}) $\xi(\tilde\epsilon)$ is determined by inverting Eq.~(\ref{tileps_Tc}).
From Eqs.~(\ref{def_tilsig_Delta}) and (\ref{tilsig_tileps_Tc}) we observe that $\tilde\sigma(\tilde\epsilon, T_c)$ is discontinuous at $\tilde\epsilon=0$,
\begin{equation}
  \lim_{\tilde\epsilon \searrow 0} \tilde\sigma(\tilde\epsilon, T_c) = 2\sigma_0 \left\{ 1 + \frac{1}{2 A^2 \ln^2 2} \right\}^{-1} \ne (2A\ln 2)^2 \sigma_0 = \tilde\sigma(\Delta, T_c). \label{tilsig_tileps_Tc_disc}
\end{equation}
The functions $\tilde\sigma(\tilde\epsilon)$ and $\sigma(\epsilon)$ are depicted in Fig.~\ref{qp_dos}.

\subsection{The FES parameters and populations}

We calculate the FES parameters using the procedure outlined in Refs. 
\cite{PhysLettA.372.5745.2008.Anghel, PhysRevE.88.042150.2013.Anghel}.
According to Eq. (\ref{inv_def_qpen}), both $\epsilon$ and $\xi$ are functions of $\tilde\epsilon$ and the occupation numbers $\{n\}$.
We express all the quantities in terms of $\tilde\epsilon$ and $\{n\}$ and we define the species by splitting the $\tilde\epsilon$ axis into small intervals $\delta\tilde\epsilon$ (see Fig. \ref{mapping}).
Each interval hosts two species, one for each type of quasiparticles, so the species corresponding to the interval $\delta\tilde\epsilon$ are denoted by $(\delta\tilde\epsilon,0)$ and $(\delta\tilde\epsilon,1)$.
The intervals are small enough, so that all the energy levels in $\delta\epsilon$ are considered degenerate, of energy $\tilde\epsilon$.
Any species $(\delta\tilde\epsilon,i)$ corresponds to two symmetric intervals $\delta\xi$ and ``$-\delta\xi$'' on the $\xi$ axis, and to an interval $\delta\epsilon$ on the $\epsilon$ axis, as shown in Fig. \ref{mapping}.
The number of single-particle states in the species $(\delta\tilde\epsilon,i)$ is $G_{\delta\tilde\epsilon} \equiv \tilde\sigma(\tilde\epsilon) \delta\tilde\epsilon = \sigma(\epsilon) \delta\epsilon = 2 \sigma_0 \delta\xi$ and the number of quasiparticles is $N_{\delta\tilde\epsilon i} \equiv G_{\delta\tilde\epsilon} n_{\tilde\epsilon i}$.
Changing $N_{\delta\tilde\epsilon i}$ is equivalent to changing $n_{\tilde\epsilon i}$.
We keep fixed the intervals $\delta\tilde\epsilon$ and this implies a variation of the corresponding intervals $\delta\xi$ and $\delta\epsilon$ with the populations $N_{\delta\tilde\epsilon i}$, according to Eqs. (\ref{inv_def_qpen}).
This leads to a variation of $G_{\delta\epsilon}$ with the population, which is the manifestation of FES.
If $\delta N_{\delta\tilde\epsilon i}$ is the (small) variation of the quasiparticle number of type $i$ in the species $(\delta\tilde\epsilon,i)$, the variation of the number of states in the species $(\delta\tilde\epsilon',j)$ in the linear approximation is $\delta G_{\epsilon'} = - \alpha_{\delta\tilde\epsilon' j; \delta\tilde\epsilon i} \delta N_{\delta\tilde\epsilon i}$, where the parameters $\alpha_{\delta\tilde\epsilon' j; \delta\tilde\epsilon i}$ are called the FES parameters. Since the number of states $G_{\delta\tilde\epsilon'}$ is the same for either type of quasiparticles, then $\alpha_{\delta\tilde\epsilon' 0; \delta\tilde\epsilon i} = \alpha_{\delta\tilde\epsilon' 1; \delta\tilde\epsilon i}$ for any $\delta\tilde\epsilon$, $\delta\tilde\epsilon'$, and $i$.

To calculate the FES parameters let us count the species in ascending order of the quasiparticle energy, like in Fig.~\ref{mapping}. The species number will be specified by a capital letter subscript $A$ or $B$ (e.g. $\delta\tilde\epsilon_A$).
Then we say that the species $(\delta\tilde\epsilon_A,i)$ corresponds to the interval $[\tilde\epsilon_A,\tilde\epsilon_{A+1})$ and the number of states in this interval is $G_{\delta\tilde\epsilon_A} = 2\sigma_0 [ |\xi(\tilde\epsilon_{A+1},\{N\})| - |\xi(\tilde\epsilon_{A}, \{N\})|]$, where by $\{N\}$ we specify explicitly the dependence of $\xi$ on the populations $N_{\delta\tilde\epsilon_A i}$.
The variation of the number of states in the species $(\delta\tilde\epsilon_A,i)$ due to the inclusion of $\delta N_{\delta\tilde\epsilon_B j}$ particles into the species $(\delta\tilde\epsilon_B, j)$ is
\begin{eqnarray}
  \delta G_{\delta\tilde\epsilon_A i} &=& 2 \sigma_0 \left[ \frac{\partial \xi(\tilde\epsilon_{A+1}, \{N\})}{\partial N_{\delta\tilde\epsilon_B j}} -  \frac{\partial \xi(\tilde\epsilon_A, \{N\})}{\partial N_{\delta\tilde\epsilon_B j}} \right] \delta N_{\delta\tilde\epsilon_B j} \nonumber \\
  &\equiv& - \delta N_{\delta\tilde\epsilon_B j} \alpha_{\delta\tilde\epsilon_A i; \delta\tilde\epsilon_B j} ,
  \label{def_alpha1}
\end{eqnarray}
where
%
%
\begin{eqnarray}
  \alpha_{\delta\tilde\epsilon_A i; \delta\tilde\epsilon_B j} &\equiv& - 2 \sigma_0 \left[ \frac{\partial \xi(\tilde\epsilon_{A+1}, \{N\})}{\partial N_{\delta\tilde\epsilon_B j}} -  \frac{\partial \xi(\tilde\epsilon_A, \{N\})}{\partial N_{\delta\tilde\epsilon_B j}} \right] . \label{def_alpha2}
\end{eqnarray}
%
%
The result (\ref{def_alpha2}) does not apply to the lowest species where we have ($A=0$)
\begin{eqnarray}
  \delta G_{\delta\tilde\epsilon_0} &=& 2 \sigma_0 \frac{\partial \xi(\tilde\epsilon_{1},\{N\})}{\partial N_{\delta\tilde\epsilon_B j}} \delta N_{\delta\tilde\epsilon_B j}
  \equiv - \alpha_{\delta\tilde\epsilon_0 i; \delta\tilde\epsilon_B j} \delta N_{\delta\tilde\epsilon_B j} . \label{def_alpha0}
\end{eqnarray}
%

To calculate the FES parameters we write
%
\begin{eqnarray}
  \frac{\partial \xi(\tilde\epsilon, \{N\})}{\partial N_{\delta\tilde\epsilon' i}} &=& \left. \frac{\partial \xi(\tilde\epsilon, \{N\})}{\partial N_{\delta\tilde\epsilon' i}} \right|_\Delta + \frac{\partial \xi(\tilde\epsilon, \{N\})}{\partial \Delta} \frac{\partial \Delta(\{N\})}{\partial N_{\delta\tilde\epsilon' i}} , \label{dxi_dN1}
\end{eqnarray}
where $[\partial \xi(\tilde\epsilon, \{N\})/\partial N_{\delta\tilde\epsilon' i}]_\Delta$ is the partial derivative of $\xi$ with respect to $N_{\delta\tilde\epsilon' i}$, when $\Delta$ is held fixed, $\partial \xi(\tilde\epsilon, \{N\})/\partial \Delta$ is the derivative of $\xi$ with respect to $\Delta$, when the populations are held fixed, and finally $\partial \Delta(\{N\})/\partial N_{\delta\tilde\epsilon' i}$ is the derivative of $\Delta$ with respect to the population $N_{\delta\tilde\epsilon' i}$.

To calculate the variation of $\Delta$ with the population on the $\tilde\epsilon$ axis we write Eq. (\ref{def_Delta2}) as
\begin{equation}
  1 = \frac{V}{2} \sum_{\tilde\epsilon} \frac{1 - n_{\tilde\epsilon 0} - n_{\tilde\epsilon 1}}{\tilde\epsilon - 2 N_\epsilon^< \cE_c/N_{ex}^2} \label{def_Delta4}
\end{equation}
from where we find
\begin{eqnarray}
  0 &=& \left[ - \frac{1}{\tilde\epsilon - 2 N_{\tilde\epsilon}^< \cE_c/N_{ex}^2}
  - \sum_{\tilde\epsilon'} \frac{1 - n_{\tilde\epsilon' 0} - n_{\tilde\epsilon' 1}}{(\tilde\epsilon' - 2 N_{\tilde\epsilon'}^< \cE_c/N_{ex}^2)^2} \left( \frac{4 N_{\tilde\epsilon'}^< \cE_c}{N_{ex}^3} - \theta (\tilde \epsilon' - \tilde\epsilon) \frac{2 \cE_c}{N_{ex}^2} \right) \right] dn_{\tilde\epsilon i} \nonumber \\
  && + \sum_{\tilde\epsilon'} \frac{1 - n_{\tilde\epsilon' 0} - n_{\tilde\epsilon' 1}}{(\tilde\epsilon' - 2 N_{\tilde\epsilon'}^< \cE_c/N_{ex}^2)^2} \frac{2 N_{\tilde\epsilon'}^<}{N^2_{ex}} \frac{\partial \cE_c}{\partial \Delta} d\Delta . \nonumber
\end{eqnarray}
Using this relation we can calculate
\begin{subequations} \label{ders_part_xi}
\begin{eqnarray}
  \frac{\partial \Delta}{\partial N_{\delta\tilde\epsilon i}} &=& \left\{ \frac{1}{\tilde\epsilon - 2 N_{\tilde\epsilon}^< \cE_c/N_{ex}^2} + \frac{2 \cE_c}{N_{ex}^2} \sum_{\tilde\epsilon'} \frac{1 - n_{\tilde\epsilon' 0} - n_{\tilde\epsilon' 1}}{(\tilde\epsilon' - 2 N_{\tilde\epsilon'}^< \cE_c/N_{ex}^2)^2} \left( \frac{2 N_{\tilde\epsilon'}^<}{N_{ex}} - \theta (\tilde\epsilon' - \tilde\epsilon) \right) \right\} \nonumber \\
  && \times \frac{N^2_{ex}}{ 4 \sigma_0 \Delta \ln\left( \Delta/\Delta_0 \right) } \left\{ \sum_{\tilde\epsilon'} \frac{1 - n_{\tilde\epsilon' 0} - n_{\tilde\epsilon' 1}}{(\tilde\epsilon' - 2 N_{\tilde\epsilon'}^< \cE_c/N_{ex}^2)^2} N_{\tilde\epsilon'}^< \right\}^{-1} ,
  \label{dDelta_dNteps}
\end{eqnarray}
where we used $\partial \cE_c/\partial \Delta = 2 \sigma_0 \Delta \ln( \Delta/\Delta_0)$, obtained from Eq. (\ref{def_Ec}).
%
%
The other terms are
\begin{eqnarray}
  \left. \frac{\partial \xi}{\partial N_{\delta\tilde\epsilon' i}} \right|_\Delta &=&
  \frac{2\cE_c}{N_{ex}^2} \frac{\tilde\epsilon - \frac{2\cE_c}{N_{ex}^2} N_{\tilde\epsilon}^<}{\sqrt{\left(\tilde\epsilon - \frac{2\cE_c}{N_{ex}^2} N_{\tilde\epsilon}^<\right)^2 - \Delta^2}} \left[ \frac{2 N_{\tilde\epsilon}^<}{N_{ex}} - \theta(\tilde\epsilon - \tilde\epsilon') \right] \quad {\rm and} \label{dxi_dN_D} \\
  \frac{\partial \xi (\tilde\epsilon, \{N\})}{\partial \Delta} &=& - \frac{\Delta}{\sqrt{\left(\tilde\epsilon - \frac{2\cE_c}{N_{ex}^2} N_\epsilon^<\right)^2 - \Delta^2}} \nonumber \\
  && \times \left[ \frac{4 \sigma_0 N_{\tilde\epsilon}^<}{N_{ex}^2} \ln \left(\frac{\Delta}{\Delta_0}\right) \left( \tilde\epsilon - \frac{2\cE_c}{N_{ex}^2} N_{\tilde\epsilon}^< \right) + 1 \right] . \label{dxi_dD}
\end{eqnarray}
\end{subequations}
Combining Eqs. (\ref{ders_part_xi}) into Eq. (\ref{dxi_dN1}) we write
\begin{subequations} \label{defs_d_n}
\begin{equation}
  \frac{\partial \xi(\tilde\epsilon_A, \{N\})}{\partial N_{\delta\tilde\epsilon_B j}} \equiv d(\tilde\epsilon_A, \{N\}) \theta(\tilde\epsilon_A - \tilde\epsilon_B)
  + e(\tilde\epsilon_A, \tilde\epsilon_B, \{N\}) , \label{dxi_dN2}
\end{equation}
where
\begin{eqnarray}
  d(\tilde\epsilon_A, \{N\}) &\equiv& - \frac{2\cE_c}{N_{ex}^2} \frac{\tilde\epsilon_A - \frac{2\cE_c}{N_{ex}^2} N_{\tilde\epsilon_A}^<}{\sqrt{\left(\tilde\epsilon_A - \frac{2\cE_c}{N_{ex}^2} N_{\tilde\epsilon_A}^<\right)^2 - \Delta^2}} , \label{def_d} \\
  e(\tilde\epsilon_A, \tilde\epsilon_B, \{N\}) &\equiv& \frac{4 \cE_c N_{\tilde\epsilon_A}^<}{N_{ex}^3} \frac{\tilde\epsilon_A - \frac{2\cE_c}{N_{ex}^2} N_{\tilde\epsilon_A}^<}{\sqrt{\left(\tilde\epsilon_A - \frac{2\cE_c}{N_{ex}^2} N_{\tilde\epsilon_A}^<\right)^2 - \Delta^2}}
  - \frac{N^2_{ex}}{ 4 \sigma_0 \ln\left( \Delta/\Delta_0 \right) } \nonumber \\
  && \times \frac{\left[ \frac{4 \sigma_0 N_{\tilde\epsilon_A}^<}{N_{ex}^2} \ln \left(\frac{\Delta}{\Delta_0}\right) \left( \tilde\epsilon_A - \frac{2\cE_c}{N_{ex}^2} N_{\tilde\epsilon_A}^< \right) + 1 \right]}{\sqrt{\left(\tilde\epsilon_A - \frac{2\cE_c}{N_{ex}^2} N_{\tilde\epsilon_A}^<\right)^2 - \Delta^2}} \nonumber \\
  && \times \left\{ \frac{1}{\tilde\epsilon_B - 2 N_{\tilde\epsilon_B}^< \cE_c/N_{ex}^2} + \frac{2 \cE_c}{N_{ex}^2} \right. \nonumber \\
  && \left. \times \sum_{\tilde\epsilon'} \frac{1 - n_{\tilde\epsilon' 0} - n_{\tilde\epsilon' 1}}{(\tilde\epsilon' - 2 N_{\tilde\epsilon'}^< \cE_c/N_{ex}^2)^2} \left( \frac{2 N_{\tilde\epsilon'}^<}{N_{ex}} - \theta (\tilde\epsilon' - \tilde\epsilon_B) \right) \right\} \nonumber \\
  && \times \left\{ \sum_{\tilde\epsilon'} \frac{1 - n_{\tilde\epsilon' 0} - n_{\tilde\epsilon' 1}}{(\tilde\epsilon' - 2 N_{\tilde\epsilon'}^< \cE_c/N_{ex}^2)^2} N_{\tilde\epsilon'}^< \right\}^{-1} .
  \label{def_n}
\end{eqnarray}
\end{subequations}
Using Eqs. (\ref{defs_d_n}) we can calculate the $\alpha$ parameters. When $A \ne B$, we write Eq. (\ref{def_alpha2}) as
\begin{eqnarray}
  \alpha_{\delta\tilde\epsilon_A i; \delta\tilde\epsilon_B j}
  &=& - 2 \sigma_0 [\xi(\tilde\epsilon_{A+1}) - \xi(\tilde\epsilon_A)] \left[\frac{\partial \xi (\tilde\epsilon, \{N\})}{\partial\tilde\epsilon}\right]^{-1} \left[ \left. \frac{\partial d(\tilde\epsilon, \{N\})}{\partial \tilde\epsilon} \right|_{\tilde\epsilon = \tilde\epsilon_A} \right. \nonumber \\
  && \left. \times \theta(A - B)
  + \left. \frac{\partial e(\tilde\epsilon, \tilde\epsilon_B, \{N\})}{\partial \tilde\epsilon} \right|_{\tilde\epsilon = \tilde\epsilon_A} \right] \nonumber \\
  &\equiv& G_{\delta \tilde\epsilon_A i} a_{\tilde\epsilon_A i; \tilde\epsilon_B j}
  , \label{def_alpha2_cont}
\end{eqnarray}
where we used $\tilde\epsilon_{A+1} - \tilde\epsilon_A = [\xi(\tilde\epsilon_{A+1}) - \xi(\tilde\epsilon_A)] \left[\partial \xi (\tilde\epsilon, \{N\})/\partial\tilde\epsilon\right]^{-1}$.
The parameters $\alpha_{\delta\tilde\epsilon_A i; \delta\tilde\epsilon_B j}$, with $A \ne B$
are \textit{extensive}, i.e. are they proportional to the dimension of the species $G_{\delta \tilde\epsilon_A i}$ on which they act \cite{JPhysA.40.F1013.2007.Anghel, EPL.87.60009.2009.Anghel}.
For the FES parameters with $A=B$ we have
\begin{eqnarray}
  \alpha_{\delta\tilde\epsilon_A i; \delta\tilde\epsilon_A j} &=& - 2 \sigma_0 d(\tilde\epsilon_A, \{N\})
  - 2 \sigma_0 [\xi(\tilde\epsilon_{A+1}) - \xi(\tilde\epsilon_A)] \frac{\partial \xi (\tilde\epsilon, \{N\})}{\partial\tilde\epsilon} \nonumber \\
  && \times \left. \frac{\partial e(\tilde\epsilon, \tilde\epsilon_B, \{N\})}{\partial \tilde\epsilon} \right|_{\tilde\epsilon = \tilde\epsilon_A}
  \equiv \tilde\alpha_{\delta\tilde\epsilon_A i; \delta\tilde\epsilon_A j} + G_{\delta \tilde\epsilon_A i} a_{\tilde\epsilon_A i; \tilde\epsilon_A j}
  . \label{def_alpha2_cont_diag}
\end{eqnarray}
%
The term $\tilde\alpha_{\delta\tilde\epsilon_A i; \delta\tilde\epsilon_A j}$ in Eq. (\ref{def_alpha2_cont_diag}) \textit{intensive} because it does not depend on the dimension $G_{\delta \tilde\epsilon_A i}$, whereas the second term $G_{\delta \tilde\epsilon_A i} a_{\tilde\epsilon_A i; \tilde\epsilon_A j}$ is extensive. All the FES parameters satisfy the rules of Ref. \cite{EPL.87.60009.2009.Anghel}, so from this point of view the formalism is consistent.

The FES equations for the two types of quasiparticles ($i,j = 0,1$) are \cite{EPL.90.10006.2010.Anghel}
\begin{eqnarray}
  \beta \tilde\epsilon_{A i} &=& \ln\frac{(1 - n_{\tilde\epsilon_A i})}{n_{\tilde\epsilon_A i}} 
  + \sum_j \sum_{B=0}^\infty \ln (1 - n_{\tilde\epsilon_B j}) \alpha_{\delta\tilde\epsilon_B j; \delta\tilde\epsilon_A i} \nonumber
\end{eqnarray}
Since neither $\tilde\epsilon$ nor $n_{\tilde\epsilon i}$ depend on the quasiparticle type (0 or 1) we write the equation above as
\begin{eqnarray}
  \beta \tilde\epsilon_A &=& \ln\frac{(1 - n_{\tilde\epsilon_A})}{n_{\tilde\epsilon_A}} 
  +2 \sum_{B=0}^\infty \ln (1 - n_{\tilde\epsilon_B}) \alpha_{\delta\tilde\epsilon_B j; \delta\tilde\epsilon_A i} , \label{Eq_FES0}
\end{eqnarray}
where $i$ and $j$ may be either 0 or 1.

\begin{figure}[t]
  \centering
  \includegraphics[width=8cm]{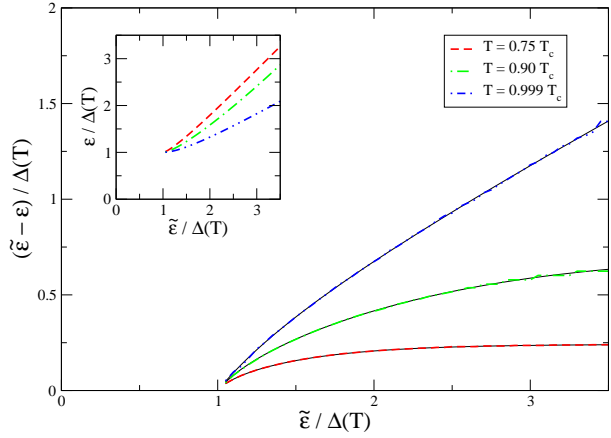}
  \caption{(Color online) 
Check of Eq. (\ref{Eq_FES_c2}) for three relevant temperatures $T=0.75,\;0.90,\;0.999$ $T_c$. Main plot: $(\tilde\epsilon-\epsilon)/\Delta(T)$ (solid black lines) and 
the second term of Eq. (\ref{Eq_FES_c2}). For each pair, the curves are identical within the numerical accuracy.
Inset: $\epsilon(\tilde\epsilon)$ scaled by $\Delta(T)$ on both axes, for each selected temperature.
}
  \label{LSH_RSH}
\end{figure}

If the formalism is consistent, then both Eqs. (\ref{pop_epsilon}) and (\ref{Eq_FES0}) should give the same populations when the quasiparticle energies are related by Eqs. (\ref{inv_def_qpen}).
To check this we use Eq. (\ref{Eq_FES0}), we transform the summations into integrals, and we write
$\alpha_{\delta\tilde\epsilon_A i; \delta\tilde\epsilon_B j} = G_{\delta \tilde\epsilon_A i} a_{\tilde\epsilon_A i; \tilde\epsilon_B j}$, with
$a_{\tilde\epsilon_A i; \tilde\epsilon_B j} = [\partial \xi (\tilde\epsilon, \{N\})/ \partial\tilde\epsilon]^{-1} [ \partial^2 \xi(\tilde\epsilon, \{N\})/ \partial \tilde\epsilon \partial N_{\delta\tilde\epsilon_B j} ]_{\tilde\epsilon = \tilde\epsilon_A}$, except for the parameters $\alpha_{\delta\tilde\epsilon_0 i; \delta\tilde\epsilon_B j}$, which are taken separately. Using integration by parts we obtain
\begin{eqnarray}
  \beta \tilde\epsilon_A &=& \ln\frac{(1 - n_{\tilde\epsilon_A})}{n_{\tilde\epsilon_A}} 
  + 2 \ln (1 - n_{\tilde\epsilon_0}) \alpha_{\tilde\epsilon_0 j; \tilde\epsilon_A i}
  + 4 \sigma_0 \int_0^\infty \ln [1 - n_{\tilde\epsilon(\xi)}] a_{\tilde\epsilon(\xi) j; \tilde\epsilon_A i} \, d\xi \nonumber \\
  %
  %
  &=& \ln\frac{(1 - n_{\tilde\epsilon_k})}{n_{\tilde\epsilon_k}} + 4 \sigma_0 \int_\Delta^\infty \frac{\partial}{\partial\tilde\epsilon}[\ln (1 - n_{\tilde\epsilon})] \frac{\partial \xi(\tilde\epsilon_{i})}{\partial N_{\tilde\epsilon_k B}} \, d\tilde\epsilon 
  %
  \label{Eq_FES_c}
\end{eqnarray}
where $i$ and $j$ are either 0 or 1.
Replacing Eq.~(\ref{pop_epsilon}) for the populations and changing the variables from $\tilde\epsilon$ to $\epsilon$ in  Eq.~(\ref{Eq_FES_c}) we get
\begin{eqnarray}
  \tilde\epsilon_A &=& 
  \epsilon(\tilde\epsilon_A) + 4 \sigma_0 \int_\Delta^\infty n_{\epsilon} \frac{\partial \epsilon [\tilde\epsilon(\epsilon), \{N\}]}{\partial N_{\tilde\epsilon_A i}} \frac{\epsilon}{\sqrt{\epsilon^2 - \Delta^2}} \, d\epsilon .
  \label{Eq_FES_c2}
\end{eqnarray}
Equation (\ref{Eq_FES_c2}) is checked numerically and found to be correct within the numerical accuracy (see Fig. \ref{LSH_RSH}).

\section{Quasiparticle-quasiparticle interaction} \label{sec_form_ext}

If we take into account as a perturbation a particle-particle interaction of the form \cite{PhysRev.108.1175.1957.Bardeen}
\begin{equation}
  \cH' \equiv \frac{1}{2} \sum_{\bk_1 \bk_2 \sigma_1 \sigma_2} \sum_{\bk_1' \bk_2' \sigma_1' \sigma_2'} V'_{\bk_1 \sigma_1 \bk_2 \sigma_2 ; \bk_1' \sigma_1' \bk_2' \sigma_2'} c_{\bk_1 \sigma_1}^\dag c_{\bk_2 \sigma_2}^\dag c_{\bk_2' \sigma_2'} c_{\bk_1' \sigma_1'}
\end{equation}
and express the electron creation and annihilation operators in terms of the $\gamma$ operators (Eqs.~\ref{def_cs}), we obtain the energies of the BCS states in the first order of perturbation theory as
\begin{eqnarray}
  E_{eff}' (\{n\}) &\equiv& 
  \cE_c + \sum_{\epsilon, i} \epsilon n_{\epsilon i}
  + \frac{1}{2} \sum_{\epsilon, \epsilon', i, j} V_{\epsilon i; \epsilon' j} n_{\epsilon i} n_{\epsilon' j} ,
  \label{HM_BCS_eff_mod}
\end{eqnarray}
instead of $E_{eff}$ of Eq. (\ref{HM_BCS_eff2}). 
By introducing new quasiparticle energies \cite{PhysRevE.88.042150.2013.Anghel}
\begin{eqnarray}
  \tilde \epsilon^{(p)}(\epsilon) &=& \epsilon + \frac{2\cE_c}{N_{ex}^2} N_{\epsilon}^< + \sum_{\epsilon' < \epsilon} \sum_{j} V_{\epsilon i; \epsilon' j} n_{\epsilon' j} . \label{def_qpen_mod}
\end{eqnarray}
we can write $E_{eff}'(\{n\}) = \sum_{\tilde\epsilon^{(p)}} \tilde\epsilon^{(p)} n_{\tilde\epsilon^{(p)}}$ as an ideal gas Hamiltonian.
In the simpler case when $V_{\epsilon i; \epsilon' j} = V_I$ is independent of $\epsilon$, $\epsilon'$, $i$, and $j$, Eqs. (\ref{HM_BCS_eff_mod}) and (\ref{def_qpen_mod}) change into 
\begin{equation}
  E_{eff}' = \left( \cE_c + \frac{V_I N_{ex}^2}{2} \right) + \sum_{\epsilon, i} \epsilon n_{\epsilon i} \quad
  {\rm and} \quad
  \tilde \epsilon^{(p)}(\epsilon) = \epsilon + \left( \frac{2\cE_c}{N_{ex}^2} + V_I \right) N_{\epsilon}^< . \label{def_qpen_mod2}
\end{equation}
From this point on the formalism may be repeated almost identically as in the previous section.

The Landau's quasiparticle energies are
\begin{equation}
  \epsilon^{(p)}(\epsilon) \equiv \frac{\partial E_{eff}'}{\partial n_\epsilon} = \epsilon + V_I N_{ex} , \label{qpen_BCS_mod}
\end{equation}
which lead to the populations
\begin{equation}
  n(\epsilon') = \frac{1}{e^{\beta\epsilon'} + 1} . \label{pop_epsilon_mod}
\end{equation}
The new FES parameters, corresponding to the energies $\tilde \epsilon^{(p)}$ (\ref{def_qpen_mod}), are related to those of Section \ref{sec_FES0} by
\begin{equation}
  \alpha^{(p)}_{\delta\tilde\epsilon^{(p)}_A i; \delta\tilde\epsilon^{(p)}_B j} = \alpha_{\delta\tilde\epsilon_A i; \delta\tilde\epsilon_B j} + 2 V_I \sigma_0 \delta_{A,B}
\end{equation}
While $\epsilon^{(p)}$ has a gap $\Delta^{(p)} \equiv \Delta + V_I N_{ex}$, the FES quasiparticle energies have the same energy gap $\Delta(T)$.
Also, similarly to Section \ref{sec_FES0}, the DOS of $\epsilon^{(p)}$ diverges at $\Delta^{(p)}$, whereas the DOS of $\tilde\epsilon^{(p)}$ never diverges, except when $T=0$ and $\tilde\epsilon^{(p)} = \Delta_0$.

At $T=0$, we have $N_{ex} = 0$ and therefore $\Delta(T=0) = \Delta_0$, like in the case $V_I=0$.
At finite temperatures the number of excitations and the gap energy are calculated from the self-consistent system of equations
\begin{subequations} \label{self_c_set}
\begin{eqnarray}
  N_{ex} &=& 4 \sigma_0 \int_\Delta^\infty \frac{1}{e^{\beta(\epsilon+V_I N_{ex})} + 1} \frac{\epsilon \, d\epsilon}{\sqrt{\epsilon^2 - \Delta^2}} , \label{self_c_Nex} \\
  \frac{1}{\sigma_0 V} &=& \int_\Delta^{\hbar\omega_c} \frac{\tanh\left[ \beta(\epsilon + V_I N_{ex})/2 \right] \, d\epsilon}{\sqrt{\epsilon^2 - \Delta^2}} . \label{self_c_Delta}
\end{eqnarray}
\end{subequations}
We observe that Eq. (\ref{self_c_Delta}) has no solution for $\Delta = 0$, if $V_I \ne 0$. 
If $V_I < 0$ the system (\ref{self_c_set}) may have multiple solutions.

\begin{figure}[t]
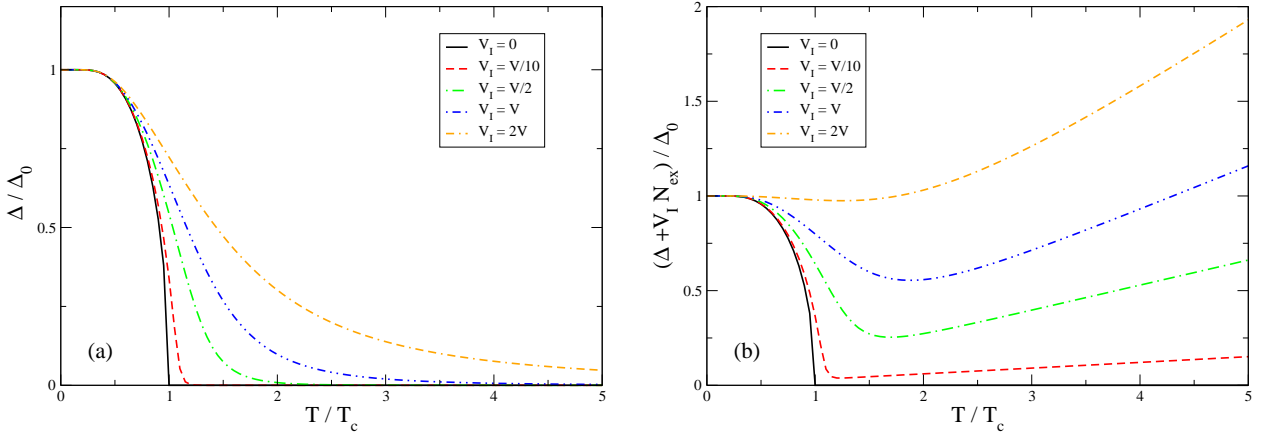

  \centering
  \includegraphics[width=8cm]{VI_Delta_T.eps}\hspace*{0.5cm}
  \includegraphics[width=8cm]{VI_Delta_plus_VINex.eps}
 \caption{The normalized gap energy of the FES (a) and Landau's (b) quasiparticle energies,  $\Delta/\Delta_0$ (Eq. \ref{def_qpen_mod}) and $(\Delta+V_I N_{ex})/\Delta_0$ (Eq. \ref{qpen_BCS_mod}), respectively, vs $T/T_c$. The black solid curve in each plot is the typical BCS gap and corresponds to $V_I=0$. $T_c$ is the BCS critical temperature. 
}
  \label{Delta_VI}
\end{figure}

In Fig. \ref{Delta_VI} we plot the gap energy for the FES (\ref{def_qpen_mod2}) and Landau's (\ref{qpen_BCS_mod}) quasiparticles energies for $V_I$ taking the values 0, $V/10$, $V/2$, $V$, and $2V$, calculated for $\sigma_0 V$ = 0.2.
For low enough $V_I$ (e.g. $V_I = V/10, V/2$) the temperature dependence of the superconducting gap $\Delta(T)$ resembles the typical BCS gap, which corresponds to $V_I=0$, but with a higher critical temperature.
On the other hand, the Landau's quasiparticles gap energy (Fig. \ref{Delta_VI}b) start to increase with the temperature when $T$ exceeds a certain value above $T_c$. This increase is due to the increase of $N_{ex}$. Since $V_I$ should be a perturbation to the typical BCS pairing Hamiltonian, the curves corresponding to $V_I=V$ and $V_I=2V$ are drawn only for comparison.
Similarly, we do not expect that the formalism should be valid for $T$ much above the BCS critical temperature, since $V_I$ is assumed to be only a perturbation to the typical pairing Hamiltonian.

\section{Conclusions} \label{conclusions}

We transformed the BCS Hamiltonian into the Hamiltonian of an ideal gas by redefining the quasiparticle energies. The new quasiparticles have the same energy gap as the BCS quasiparticles, but their density of states is finite over the whole quasiparticle spectrum. The ideal gas thus obtained obeys fractional exclusion statistics (FES). We calculated the FES parameters and showed that this description is thermodynamically equivalent to the standard BCS description, in the sense that at given temperature it leaves unchanged the populations of the quasiparticle states and all the thermodynamic quantities.

We introduced an electron-electron perturbation interaction. In the basis of the BCS state functions this perturbation is expressed as a quasiparticle-quasiparticle interaction which stabilizes the condensate increasing the critical temperature.
We determined the quasiparticle energies and the gap energy in the presence of the perturbation.

\section{Acknowledgments}

This work has been financially supported by CNCSIS-UEFISCDI (project IDEI 114/2011) and ANCS (project PN-09370102). Travel support from Romania-JINR Collaboration grants 4436-3-2015/2017, 4342-3-2014/2015, and Titeica-Markov program is gratefully acknowledged.


\end{document}